# Cognitive Networks and Performance Drive fMRI-Based State Classification Using DNN Models

Murat Kucukosmanoglu[1], Javier O. Garcia[2], Justin Brooks[1,3,4], Kanika Bansal[2,3,*]

[1] D-Prime LLC, McLean, VA 22101 USA
[2] Humans in Complex Systems, US DEVCOM Army Research Laboratory, Aberdeen Proving Ground, MD 21005 USA
[3] Department of Computer Science and Electrical Engineering, University of Maryland, Baltimore County, Baltimore, MD 21250 USA
[4] Tanzen Medical Inc., Baltimore, MD, USA

* Corresponding author email: phy.kanika@gmail.com

## Abstract

Deep neural network (DNN) models have demonstrated impressive performance in various domains, yet their application in cognitive neuroscience is limited due to their lack of interpretability. In this study we employ two structurally different and complementary DNN-based models, a one-dimensional convolutional neural network (1D-CNN) and a bidirectional long short-term memory network (BiLSTM), to classify individual cognitive states from fMRI BOLD data, with a focus on understanding the cognitive underpinnings of the classification decisions. We show that despite the architectural differences, both models consistently produce a robust relationship between prediction accuracy and individual cognitive performance, such that low performance leads to poor prediction accuracy. To achieve model explainability, we used permutation techniques to calculate feature importance, allowing us to identify the most critical brain regions influencing model predictions. Across models, we found the dominance of visual networks, suggesting that task-driven state differences are primarily encoded in visual processing. Attention and control networks also showed relatively high importance, however, default mode and temporal-parietal networks demonstrated negligible contribution in differentiating cognitive states. Additionally, we observed individual trait-based effects and subtle model-specific differences, such that 1D-CNN showed slightly better overall performance, while BiLSTM showed better sensitivity for individual behavior; these initial findings require further research and robustness testing to be fully established. Our work underscores the importance of explainable DNN models in uncovering the neural mechanisms underlying cognitive state transitions, providing a foundation for future work in this domain.

## 1. Introduction

Deep neural networks (DNNs) have been integrated with functional-MRI data for various applications, such as distinguishing pathological dynamics from healthy ones (Deshpande et al., 2024; Quaak et al., 2021), understanding decision making (Fintz et al., 2022), and classifying stress states (Shermadurai and Thiyagarajan, 2023). In certain instances, these challenges have been addressed with notable success. For example, the identification of disease states, including autism spectrum disorder (Wang et al., 2019), schizophrenia (Chyzhyk et al., 2015; Patel et al., 2016; Qureshi et al., 2019; Srinivasagopalan et al., 2019), and ADHD (Deshpande et al., 2015), compared to healthy individuals, has been demonstrated with high accuracy (exceeding 90%). While these studies have paved the way for incorporating DNN-based models in neuroscience, many findings, beyond study-specific limitations, suffer from the issues of interpretability and generalizability (Quaak et al., 2021). Indeed, these challenges represent the primary obstacles in utilizing DNN-based models to improve our understanding of human cognition, which could lead to novel technological and clinical advancements.

In particular, recent studies using DNN-based state classifiers, on the one hand demonstrate impressive performance and high classification accuracy (Gupta et al., 2022; Meng and Ge, 2022; Meszlényi et al., 2017), but on the other suffer from the lack of interpretability and insights into the underlying mechanisms driving classification decisions (Thomas et al., 2022). A more transparent insight into these mechanisms will not only improve our understanding of the brain's large-scale cognitive function, but may also provide ways in which individual variability, behavioral traits, and heterogeneities and commonalities across disorders or dysfunction can be understood and addressed (Rungratsameetaweemana et al., 2022; Segal et al., 2023). Researchers are actively developing methods to enhance the interpretability of deep learning models applied to fMRI data, such as layer-wise relevance propagation or attention mechanisms (Čík et al., 2021; Li et al., 2022, 2019). Besides these efforts, a substantial amount of work focused on achieving high performance accuracy, leading to limited generalizability and neural relevance. Importantly, how the performance of DNN models relate to "individual states" is an open question.

In this study, we developed two DNN models using common algorithms to distinguish cognitive states from fMRI-derived BOLD signals, where individuals engaged in different cognitive demands. We aimed to gain a deeper understanding of the neural underpinnings of the classification process. Specifically, we investigated how individual performance on cognitive tasks affects classification accuracy and, to enhance explainability, we leveraged permutation techniques to assess the dominant brain networks driving the classification. Permutation techniques measure the impact of a feature by evaluating how much the model's performance declines when the feature's values are randomly shuffled (Breiman, 2001; Chamma et al., 2023). Similar to the model-agnostic approach of permutation variable importance (Covert et al., 2021; Janitza et al., 2018), this method allowed us to identify the most critical brain regions influencing model predictions, thereby improving the reliability and interpretability of our models for fMRI data analysis. Additionally, we examined how these relationships depend on model architecture.

We used one-dimensional convolutional neural network (1D-CNN) and a bidirectional long short-term memory network (BiLSTM) model, structurally designed to target distinct features of the spatial-temporal signals, with a rich neuroimaging dataset where participants performed a variety of classic cognitive tasks (Nakuci et al., 2023), targeting processes including but not limited to attention, vigilance, and working memory, in a longitudinal fashion, across a period of sixteen weeks. The participants showed behavioral variability crucial to assess the relationship between model performance and individual performance. Through our findings we discuss robust consistencies, accompanied by subtle differences, in the cognitive underpinnings of cognitive state classification using two popular DNN architectures. This work lays a foundation to compare and understand the performance of DNN models, as driven by underlying cognitive factors. Our findings provide ways in which new insights can be achieved on the various neurological processes, as captured by the large-scale neuroimaging data.

## 2. Material and methods

### 2.1 Experimental design

The experimental protocol included biweekly sessions, up to 16 weeks (8 sessions), during which physiological data was collected, including fMRI, while participants (N= 56, age 22.2 ± 3, 58% female, from the greater Santa Barbara area) engaged in a variety of tasks that are well established in literature to probe fundamental cognitive processes such as vigilance, working memory, and attention (Figure 1A). These tasks included psychomotor vigilance task (PVT), visual working memory task (VWM), dot probe task (DOT), modular math task (MOD), and dynamic attention task (DYN). PVT is a classic task to assess vigilance where participants were asked to respond to the appearance of a stimulus as fast as they can. In VWM participants were presented with squares of different colors for 150 milliseconds and after a delay period of 1180 milliseconds, were asked to recall whether the new presentation was the same or different from the previously presented stimuli. DOT, a standard task designed to probe selective attention, required the participants to quickly choose between visual cues on the opposite sides (images with positive and negative valence). DYN included multiple object tracking. For MOD, participants were presented with a modular arithmetic expression with remainder after dividing two integers and were asked if the expression was correct. Tasks contained multiple trials where difficulty levels were randomized. In addition to these tasks, five minutes resting state data was also collected in every session. Note that every participant was not able to complete all the eight sessions initially planned, and some participants left the experiment early. Therefore, the number of sessions varied between 1 to 8 across participants (n = 23 for only 1 session, n = 23 for all 8 sessions). However, every session included five task states and one resting state. All methods and procedures in the present study were approved by both of the Institutional Review Boards at the University of California, Santa Barbara and the U.S. DEVCOM Army Research Laboratory. All participants provided written informed consent to participate.

## 2.2 Behavioral performance

Task performance was included from PVT, VWM, and MOD tasks for each session. PVT performance was measured as average reaction time across trials within a session. VWM and MOD task performance was measured as the accuracy, i.e., proportion of correct trials in a session. To assess the impact of task performance on model prediction accuracy, we calculated a representative performance score for each session. First, for every task we divided all the sessions across individuals into four quartiles based on performance scores such that quartile 1 included worst performance sessions and quartile 4 included best performance sessions. Then, we averaged quartiles across tasks to gauge how well an individual performed during a session, on these tasks in general. A high average (closer to 4) for a session represents consistent high task performance on all the tasks during that session, a low average score (close to 1) represents consistently poor performance on all the tasks during that session, and scores in between represent partially good/bad task performance. To obtain categorization, we calculated quartiles of this average as a representative behavioral performance metric, which we call *effective behavior quartile*. Note that an individual can fall into different effective behavior quartiles for different measurement sessions across the length of the experiment.

## 2.3 fMRI data acquisition and preprocessing

Details of the fMRI data acquisition and preprocessing can be found in (Nakuci et al., 2023). In brief, f-MRI data was acquired on a 3 T Siemens Prisma MRI using an echo-planar imaging (EPI) sequence (voxel size: 3 × 3 × 3 mm, repetition time (TR) = 910 ms). fMRI BOLD (Blood Oxygenation Level Dependent) images were preprocessed using Advanced Normalization Tools (ANTs) (Avants et al., 2011). Head motion correction as well as physiological artifacts (including respiration and cardiac cycle) removal was performed using standard methods (Nakuci et al., 2023). Finally, the voxel-wise fMRI volumes were transformed into MNI space and BOLD time-series were extracted for 214 brain regions; 200 cortical regions defined by the Schaefer 200 atlas (Schaefer et al., 2018) and 14 subcortical regions from the Harvard–Oxford atlas (Makris et al., 2006). As the atlases are in MNI coordinate space, voxels within each labeled region of the atlases were simply averaged, and time series were extracted for the following analyses. For each individual and session, the regional time series was z-scored to standardize the signal.

The fMRI time series data varied in signal length across tasks. Given the considerable variability in mean sequence length, which ranged approximately from 290 to 800 time steps, and class imbalance across the dataset, we segmented the data into fixed-length segments (Figure 1B). This segmentation was carried out to facilitate subsequent model training and analysis (Figure 1C), particularly for models such as BiLSTM (Schuster and Paliwal, 1997) and 1D-CNN, which benefit from uniform input lengths during training.

Each segment contained 277 time points (Figure 1D), a sequence length carefully selected to ensure consistency in input size for subsequent model training, while also maintaining the cognitive relevance and adequate sample size. In cases where segments had fewer than 277 time points but more than 267 points (only 2%), zero-padding was applied to achieve a uniform sequence length for batch training. Segments shorter than 267 were not included in the samples. Figure 1D shows the number of segments (samples) for each task state across the entire dataset after segmentation. In total we included 3497 samples across individuals and tasks. Different samples derived from the same session for an individual were assigned the individual effective behavior quartile for that particular session day. This strategic preprocessing approach minimized data loss and ensured robustness in subsequent analyses. By standardizing sequence lengths, we optimized model training efficiency and reduced potential biases introduced by variable input sizes.

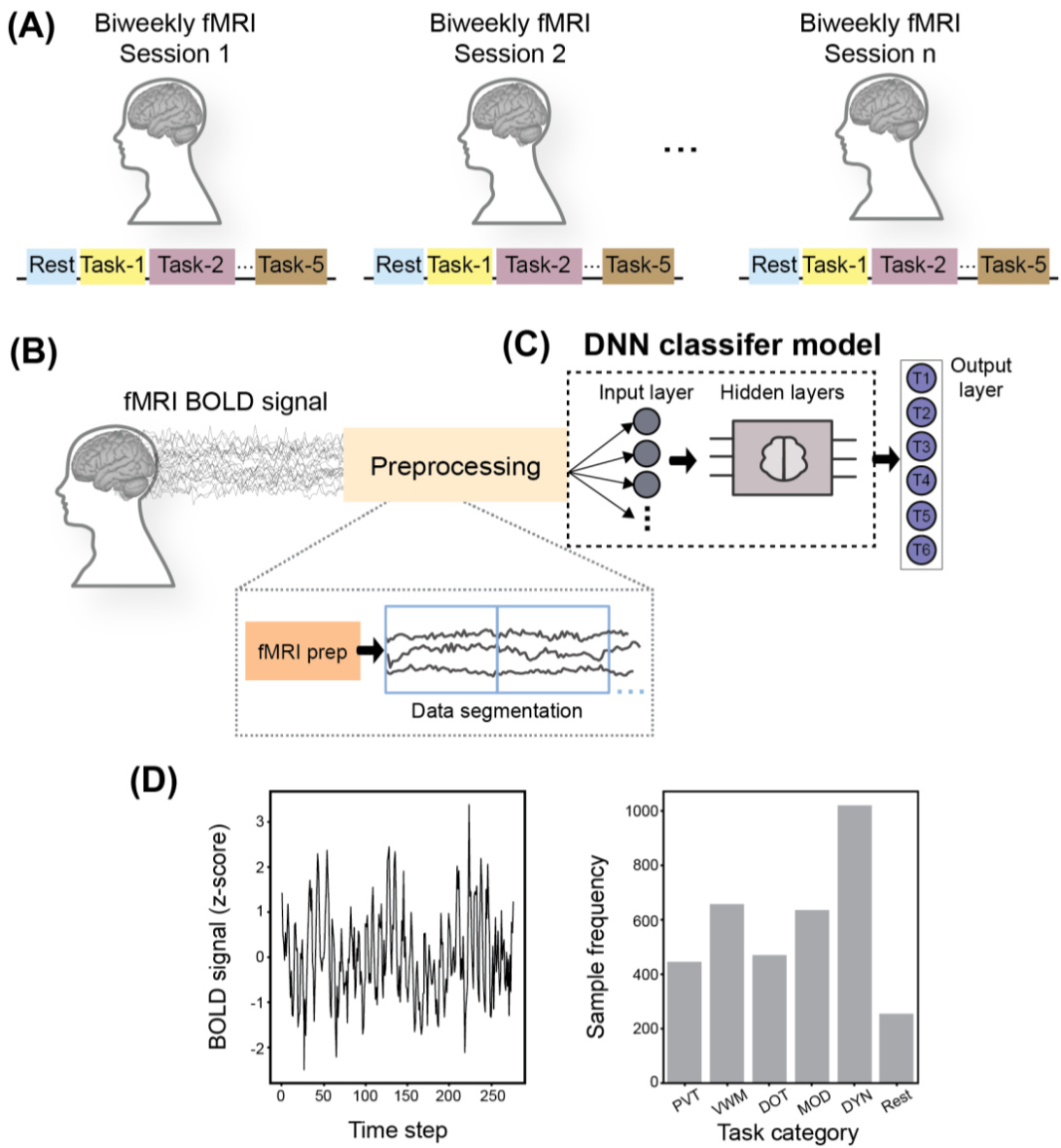

**Figure 1**: Study design overview. (A) The experimental protocol included biweekly sessions during which fMRI data was collected while participants engaged in various cognitive tasks. (B) Given the variability in task durations, after standard fMRI data preprocessing, the signal was segmented into smaller sample durations across tasks, sessions, and participants, in order to homogenize the sample size for model training and minimize data loss. (C) Two DNN models (detailed in the next figure) were utilized independently to perform model training and task classification. (D) Example data segment along with sample frequency across tasks.

## 2.4 DNN models for classification of task states

To classify different task states, we utilized 1D-CNN and BiLSTM models as they are exceptionally well-suited for analyzing fMRI data, particularly due to their proficiency in handling time-series information. 1D-CNNs excel at capturing local temporal patterns, much like their effectiveness in computer vision tasks where CNN layers demonstrate remarkable prowess in detecting features. This makes them apt for discerning intricate regional relationships within the brain. On the other hand, BiLSTM networks possess a remarkable ability to uncover temporal dependencies within sequential data. This capability is especially vital for processing the dynamic nature of brain activity over time, a crucial aspect in understanding brain functional relationships. By harnessing the strengths of 1D-CNNs and BiLSTMs independently, we aimed to explore brain functional activity across spatial and temporal domains. This approach allows us to delve into the intricate interplay between different brain regions over time. In the following we provide further details of the models utilized.

**1D-CNN:** We designed a 1D-CNN architecture specifically for analyzing and classifying temporal patterns within fMRI time-series data from 214 regions (Figure 2A). The CNN model utilizes convolutional layers to extract local temporal features from the input data. These layers perform convolutional operations, which involve sliding kernels (small filters) over the input signal to detect patterns and features. Our model architecture takes the form of a sequential CNN model, comprising multiple layers. Each section of the model consists of a convolutional layer, followed by a dropout layer, and then a max-pooling layer. This structure allows the model to progressively learn and generalize temporal patterns in the fMRI time-series data. The convolutional layers in the model use the ReLU (Rectified Linear Unit) activation function, which helps in introducing non-linearity to the model, enabling it to learn complex patterns.

To ensure robust and efficient training, our 1D-CNN model incorporates several key components: **(1) Dropout regularization**: applied after convolutional layers to mitigate overfitting by randomly deactivating neurons during training. This technique helps in enhancing the generalization capability of the model. In our model, the dropout ratio is set to 0.4. This specific ratio was chosen based on empirical evidence suggesting that a 40% dropout rate effectively balances the trade-off between overfitting and underfitting. **(2) Early stopping**: implemented to monitor validation loss and halt training when the model's performance plateaus. This prevents overfitting and ensures that the model retains its ability to generalize well to unseen data. By setting the patience parameter to 10, the model is allowed to continue training for 10 more epochs after the validation loss stops improving. **(3) Learning rate reduction**: employed to adapt the optimization process, enhancing training stability and convergence. This ensures that the learning process is efficient and that the model reaches optimal performance. **(4) Grid Search:** conducted a comprehensive grid search to find the best hyperparameters. This process involved systematically varying parameters such as the number of filters, kernel size, dropout rate, batch size, and learning rate to identify the combination that resulted in the highest validation accuracy, while also preventing overfitting to the training dataset.

The final layer of the model is a dense (fully connected) layer that outputs probabilities for each task. This layer uses a softmax activation function to convert the raw output scores into probabilities, indicating the likelihood that each input time-series belongs to a particular class. The highest probability is taken as the final prediction, representing the class to which the input most likely belongs. Below is Table 1, describing the layers in the 1D-CNN model along with the output shapes and parameter counts.

**Table 1: Summary of 1D-CNN architecture for sequence processing.** The output of each layer is the input for the next layer.

| Layer (type) | Output shape | Parameter # |
|---|---|---|
| InputLayer | (None, 277, 214) | 0 |
| Conv1D (ReLU) | (None, 277, 64) | 41,152 |
| Dropout (0.4) | (None, 277, 64) | 0 |
| MaxPooling1D | (None, 138, 64) | 0 |
| Conv1D (ReLU) | (None, 138, 128) | 24,704 |
| Dropout (0.4) | (None, 138, 128) | 0 |
| MaxPooling1D | (None, 69, 128) | 0 |
| Flatten | (None, 8,832) | 0 |
| Fully Connected Layer (ReLU) | (None, 128) | 1,130,624 |
| Dropout (0.4) | (None, 128) | 0 |
| Output Layer (SoftMax) | (None, 6) | 774 |

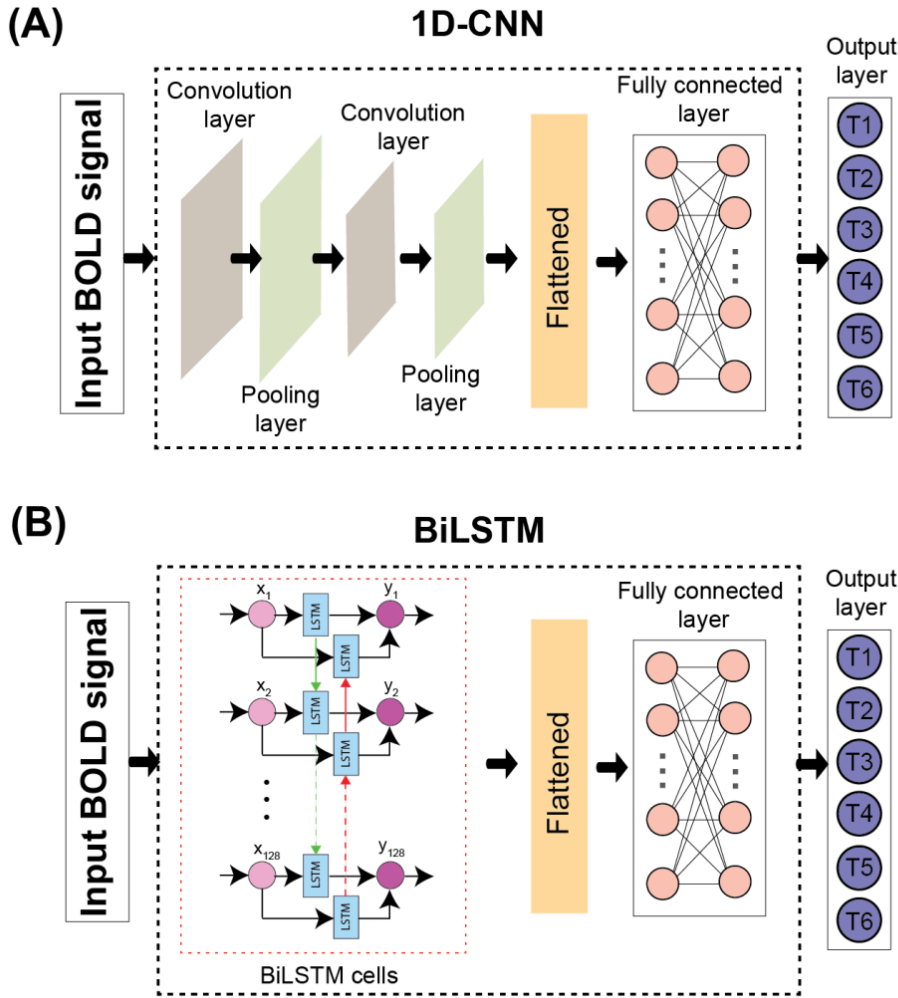

**Figure 2**: DNN models architecture. (A) 1D-CNN architecture employed for fMRI signal analysis. Dropout layers are not explicitly shown, as they are not the main functional layers but serve specific purposes such as regularization and reshaping the data. (B) BiLSTM architecture employed for fMRI signal analysis.

**BiLSTM**: The BiLSTM model processes input data bidirectionally, employing two sets of LSTM layers (Figure 2B). One set operates in the forward direction, analyzing input from the beginning to the end, while the other set operates in the backward direction, processing input from the end to the beginning. This bidirectional approach enables the model to capture information from both past and future contexts, enhancing its ability to discern longer temporal dependencies within the data. Input data for the BiLSTM model consists of fMRI time-series data, representing measurements taken over time at various brain locations (214 regions). Each region's time-series data is treated as a separate input feature. The model processes the multidimensional input data in a sequential manner, considering all regions at each time step. This means that at each time step, the BiLSTM simultaneously takes in data from all 214 regions, capturing temporal dependencies and patterns both within and across different brain regions.

Within each LSTM cell, four types of gating mechanisms are employed: Input Gate (i): determines the significance of new input data. It decides which parts of the input data should be passed on to the cell state. This gate uses the sigmoid activation function. Forget Gate (f): determines the significance of retaining or forgetting previous information stored in the cell state. It decides which information from the previous cell state should be forgotten or retained. This gate uses the sigmoid activation function. Cell Update Gate (g): updates the cell state by combining the new input data with the previous cell state. It computes the new candidate values to be added to the cell state. This gate uses the tanh activation function. Output Gate (o): determines the significance of outputting information from the cell state to the next LSTM layer or the final output. It controls which parts of the cell state should be exposed as the output. This gate uses the sigmoid activation function. These gating mechanisms enable the model to retain and update relevant information over time, facilitating the learning of meaningful representations or patterns within the data.

To ensure robust and efficient training, following key factors were incorporated: **(1) Dropout regularization**: dropout regularization, with a ratio of 0.5, is applied to mitigate overfitting by randomly deactivating neurons during training, enhancing the model's generalization capability. **(2) Early stopping**: setting the patience parameter to 10 allows the model to continue training for 10 more epochs after the validation loss stops improving, preventing unnecessary training and overfitting. **(3) Learning rate reduction**: this ensures that the learning process is efficient and that the model reaches optimal performance. **(4) Grid Search:** conducted a grid search to find the best hyperparameters by varying the number of LSTM units, dropout rate, batch size, and learning rate.

Each parameter has a temporal aspect. For example, the patience parameter of 10 ensures the model is not influenced by short-term fluctuations, focusing instead on stable patterns over a longer training period. This allows for the identification of robust temporal features underlying specific cognitive processes. Table 2 describes the layers in the BiLSTM model we constructed along with the output shapes and parameter counts.

**Table 2: Summary of BiLSTM architecture for sequence processing.** The output of each layer is the input for the next layer.

| Layer (type) | Output shape | Parameter # |
|---|---|---|
| InputLayer | (None, 277, 214) | 0 |
| Bidirectional (tanh/sigmoid) | (None, 277, 128) | 142848 |
| Flatten | (None, 35456) | 0 |
| Dropout (0.5) | (None, 35456) | 0 |
| Fully Connected Layer (ReLU) | (None, 64) | 2269248 |
| Dropout (0.5) | (None, 64) | 0 |
| Output Layer (SoftMax) | (None, 6) | 390 |

## 2.5 Model training and evaluation

For the training and testing process, we ensured the separation of validation data from the training and test data sets. The allocation was approximately 70% for training, 10% for validation, and 20% for testing. Importantly, we ensured that data from the same subjects performing the same tasks was kept separate across the training, validation, and testing sets. This approach allowed for a reliable evaluation of the model's performance. This distribution was crucial to ensure an unbiased evaluation of the model's performance.

We assessed model performance using key metrics such as overall accuracy, precision, recall, and the F1-score. Overall accuracy, defined by the percentage of correctly predicted observations across all tasks (Equation 1), provides a broad view of the model's effectiveness. Precision (Equation 2), recall (Equation 3), and the F1-score (Equation 4) were evaluated to examine the model's capabilities in handling task-specific predictions. Precision, the ratio of true positive predictions to all positive predictions made by the model, emphasizes its ability to minimize false positives. Conversely, recall, which measures the proportion of actual positives correctly identified, highlights the model's effectiveness in capturing all relevant instances. The F1-score, a harmonic mean of precision and recall, offers a single metric that balances the concerns of both precision and recall. These equations collectively describe and quantify the performance of the model in various aspects:

$$Accuracy = \frac{Number\ of\ Correct\ Predictions}{Total\ Number\ of\ Observations}$$

(1)

$$Precision = \frac{True\ Positives}{True\ Positive + False\ Positives}$$

(2)

$$Recall = \frac{True\ Positives}{True\ Positive + False\ Negatives}$$

(3)

$$F1 = 2\frac{(Precision\ x\ Recall)}{(Precision + Recall)}$$

(4)

Further, we conducted a feature importance assessment focusing on the influence of various brain regions on task classification. This involved a permutation importance technique where the BOLD signals of selected brain regions (functional networks derived from the Schaefer atlas) were replaced with random noise, generated with a mean of 0 and scaled such that roughly 95% of its values fell within the range of [−2, 2] to mimic the distribution of the experimental data. Similar to the consensus approach with permutation steps discussed in (Deshpande et al., 2024), this method allowed us to identify the most critical brain regions influencing model predictions, thereby improving the reliability and interpretability of our DNN models for fMRI data analysis

To assess overall feature importance across tasks, we analyzed the drop in overall accuracy of the mode after permuting the selected regions. For task-specific feature importance, we assessed the drop in the F1-score after permuting regions, which is the optimal metric for evaluating class-specific performance. We interpreted a greater decline in accuracy and F1 score as indicative of a feature's significant role in task classification. This methodology allowed us to pinpoint the brain regions most critical to the classification of cognitive tasks, thereby offering deeper insights into the cognitive underpinnings of the model performance.

## 3. Results

### 3.1 Model performance

Figure 3 shows the training and validation metrics for the 1D-CNN model, while Figure 4 illustrates the training and validation metrics for the BiLSTM Model. Both models have been selected with the best hyperparameters.

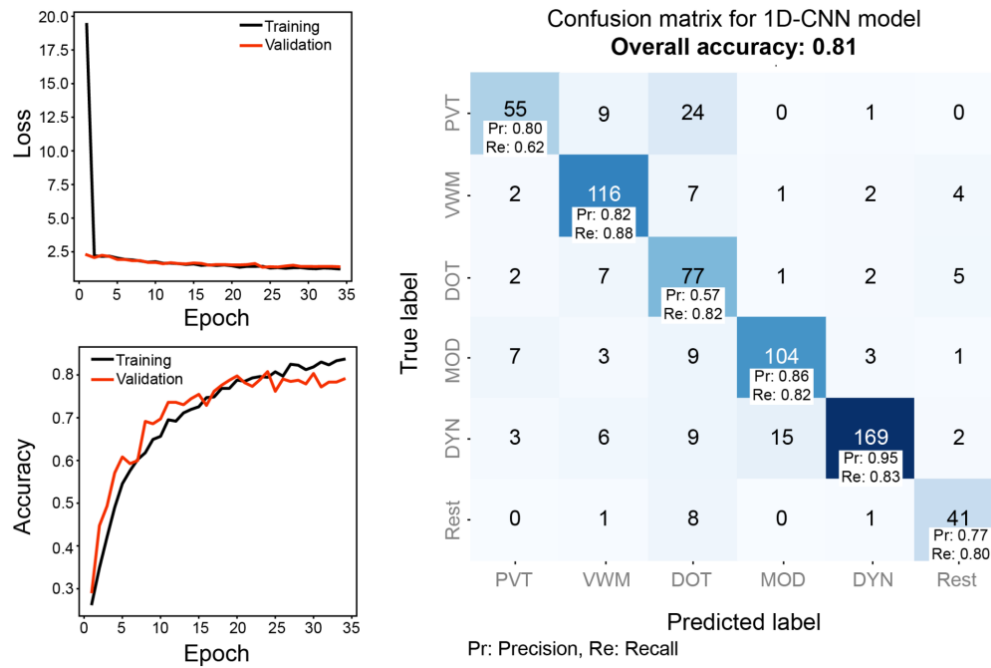

**Figure 3:** Training and validation metrics along with confusion matrix for 1D-CNN model. Upper left: Training and validation loss over epochs demonstrating the model's convergence. Lower left: Training and validation accuracy over epochs indicating the model's learning progress.

Right: Confusion matrix displaying the model's classification performance on test data across different classes (tasks). Heatmap cells represent the count of true positive predictions per task.

For the 1D-CNN model (Figure 3), we achieved an overall accuracy of 0.81. The precision and recall for specific tasks were PVT: Precision = 0.80, Recall = 0.62, VWM: Precision = 0.82, Recall = 0.88, DOT: Precision = 0.57, Recall = 0.82, MOD: Precision = 0.86, Recall = 0.82, DYN: Precision = 0.95, Recall = 0.83, and Rest: Precision = 0.77, Recall = 0.80.

For the BiLSTM model (Figure 4), we achieved an overall accuracy of 0.78. The precision and recall for specific tasks were PVT: Precision = 0.57, Recall = 0.55, VWM: Precision = 0.86, Recall = 0.80, DOT: Precision = 0.62, Recall = 0.68, MOD: Precision = 0.81, Recall = 0.86, DYN: Precision = 0.86, Recall = 0.83, and Rest: Precision = 0.85, Recall = 0.88.

We found DYN to be the best predicted task across both models. VWM also had high classification accuracy consistently across both models. Notably, across models, the classification accuracy for DOT was relatively lower with lower recall and precision scores compared to others, suggesting challenges in classification for this specific class. Additionally, PVT showed the lowest accuracy for BiLSTM. In general, we found that the 1D-CNN model outperformed the BiLSTM model by 3% in overall accuracy, indicating that the 1D-CNN model may be more effective for this type of data and classification tasks. The 1D-CNN model also exhibited higher precision and recall across most tasks compared to the BiLSTM model, highlighting its superior performance in this context.

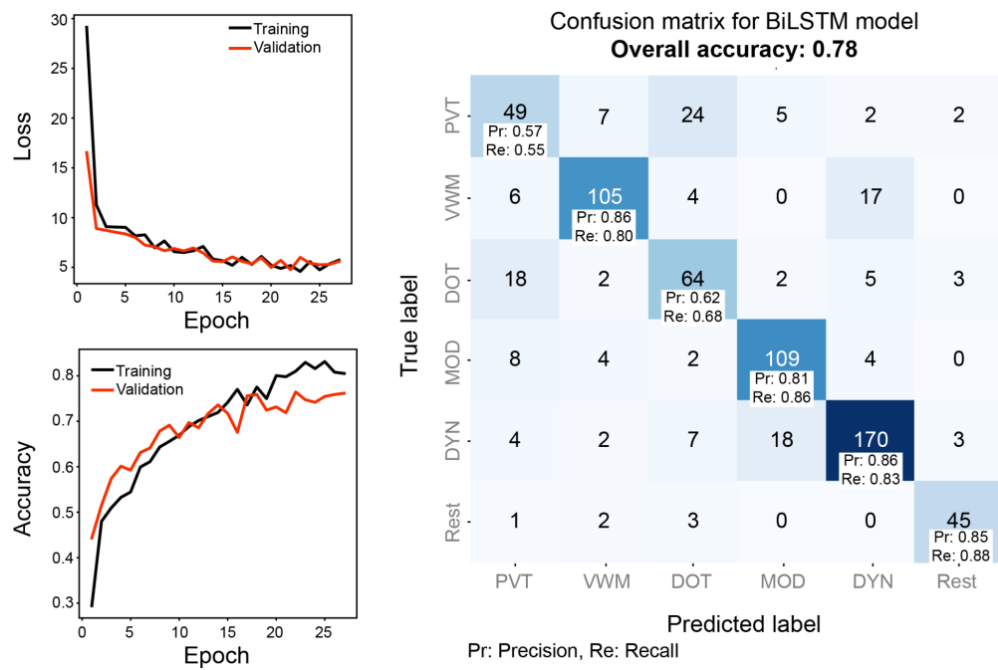

**Figure 4**: Training and validation metrics along with confusion matrix for BiLSTM model. Upper left: Training and validation loss over epochs demonstrating the model's convergence. Lower left: Training and validation accuracy over epochs indicating the model's learning progress. Right: Confusion matrix displaying the model's classification performance on test data across different classes (tasks). Heatmap cells represent the count of true positive predictions per task.

## 3.2 Fraction of incorrect predictions by individuals

To further understand the nuances of model prediction errors and its interactions with individual performance and variability, for each individual, we calculated the fraction of incorrect predictions across all the tasks and compared this ratio for two models (Figure 5A). Note that given the variability in number of sessions and segments across individuals, to reduce any bias in fractions due to the low number of sessions, in this analysis we included the individuals for whom the total number of segments available for classification were 10 or more, with the range being from 4 to 29.

We observed a strong positive correlation (Pearson's $r = 0.69$, $p$-value = 0.00001) in fraction of incorrect predictions between BiLSTM and CNN models (Figure 5B) suggesting that despite their architectural differences, the models exhibit comparable performance. This consistency in predicted outcome across individuals suggests a common lack of task specific signatures during certain individual states (sessions).

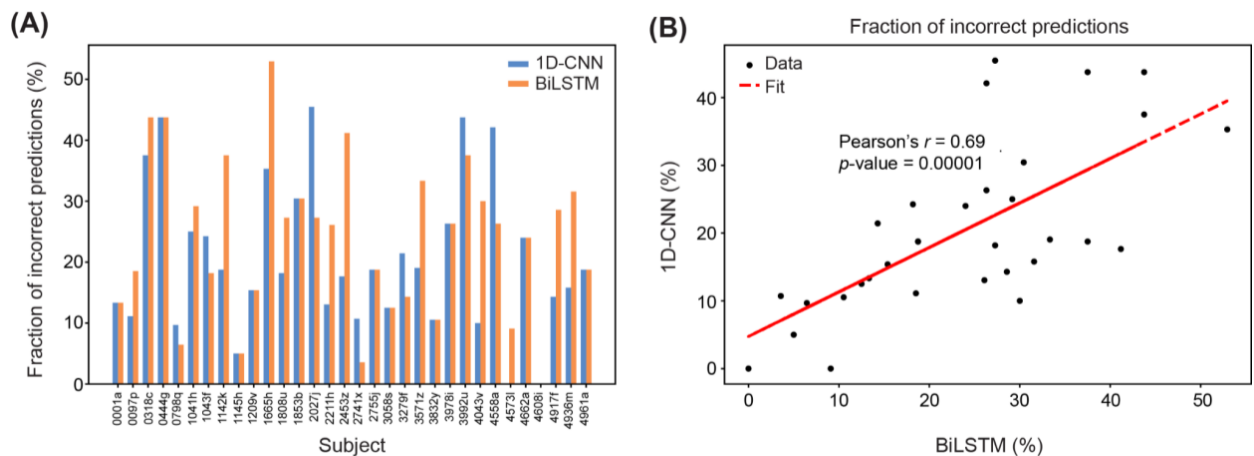

**Figure 5:** (A) The fraction of incorrect predictions for each subject in both 1D-CNN (blue) and BiLSTM models. Bars indicate the percentage of incorrect predictions. (B) Scatter plot for the fraction of incorrect predictions per subject for BiLSTM versus 1D-CNN model, with a regression line (red dashed) fitted to the data points.

Specifically, certain subjects, such as 1665h, 2027j, and 3992u, showed a higher number of incorrect predictions compared to others (Figure 5A). This could stem from reduced engagement with the task. We hypothesized that the frequency of inaccurate predictions may indicate a diminished specificity in the neural dynamics essential for precise model classification, also resulting in decreased task performance. In our next analysis we sought to investigate this hypothesis.

## 3.3 Relationship between model performance and behavioral performance

We categorized the predictions into two groups: incorrect and correct, based on the test dataset. In Figure 6A-B we plot the Effective Behavior Quartile (EBQ, Materials and methods) for incorrect and correct groups for 1D-CNN and BiLSTM model respectively. We observed a clear significant difference of performance between groups (on Welch's t-test), with correct predictions showing better performance scores (higher quartile). Specifically, for the 1D-CNN model, $t$-statistic = 2.447, $p$-value = 0.01520 and for the BiLSTM model, $t$-statistic = 3.773, $p$-value = 0.00020, indicating a significant difference in performance between the correct and incorrect prediction groups.

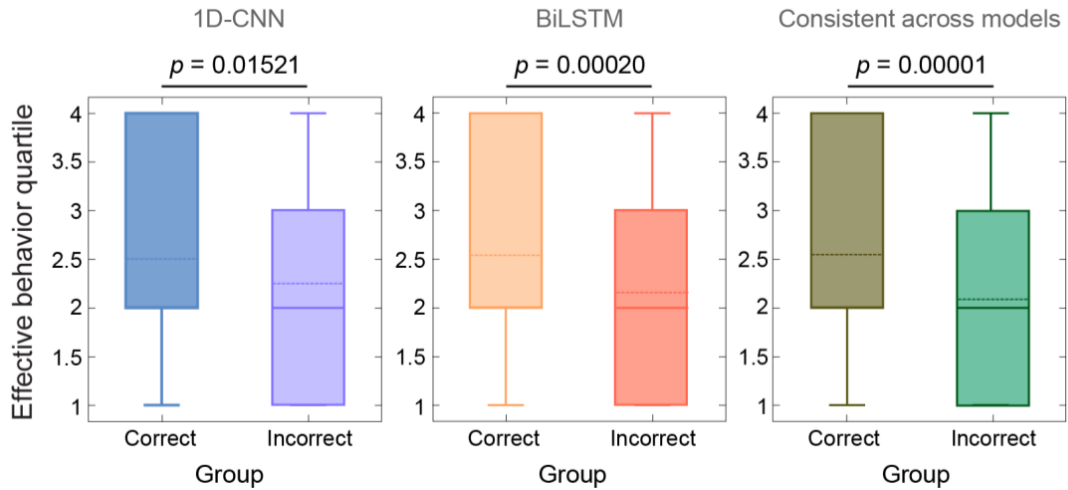

**Figure 6:** Effective behavior quartiles for correct and incorrect prediction groups for 1D-CNN model (left), for BiLSTM model (center), and for consistently predicted correct or incorrect across both models (right). The solid line within each box indicates the median, and the dashed line represents the mean. To test for significant differences, we performed Welch's t-test and the obtained $p$-values are included here.

Further, we also separated the instances where both the models predicted either correct or incorrect outcomes consistently. In Figure 6C we compare these consistent groups and observe an even stronger effect with $t = 4.609$ and $p$-value = 0.00001. Indicating a robust relationship between incorrect model prediction and low individual task performance. This suggests that biological biomarkers in the fMRI signal are more pronounced in cases of better performance.

### 3.3.1 Individual differences in prediction outcome

To delve deeper into this analysis and determine if some individuals were more likely to be predicted incorrectly due to specific performance traits, we divided individuals into two categories based on the model's predictions. Using the median value of the ratio of incorrect prediction, shown in Figure 5, we classified subjects with a lower ratio of incorrect predictions as *well-predicted* and those exceeding the median as *ill-predicted*. This process avoided duplicating subjects across groups. As shown in Figure 7A-B, for these two groups we compared the effective behavior quartiles and observed a significant difference such that well-predicted individuals showed better task performance (1D-CNN: $t = 5.7250$, $p = 2 \times 10^{-8}$; and BiLSTM: $t = 3.1179$, $p = 0.0019$).

Within well-predicted and ill-predicted groups, we further investigated the EBQ by separating correct and incorrect predictions. For the ill-predicted group, we observed a significantly higher EBQ for correct predictions in both the models (1D-CNN: $t = 2.32$, $p = 0.02$; BiLSTM: $t = 2.12$, $p = 0.03$). For well-predicted groups however, EBQ were comparable across incorrect and correct predictions ($p > 0.05$) for 1D-CNN, whereas, correct predictions lead to higher EBQ for BiLSTM ($t = 2.21$, $p = 0.03$). This key model difference likely indicates better sensitivity for individual traits for BiLSTM architecture and a stronger relationship between model prediction and behavioral performance. This might make BiLSTM a potentially better choice for applications requiring detailed behavioral insights.

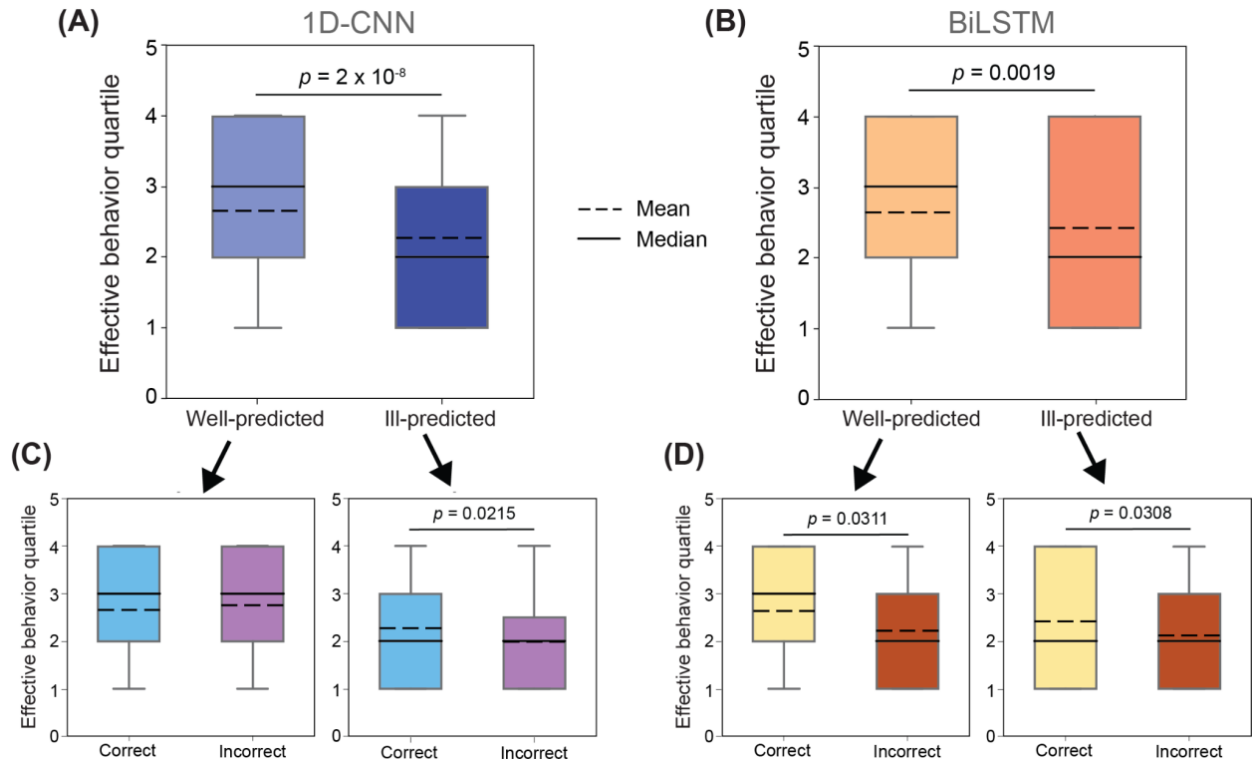

**Figure 7:** Comparison of effective behavior quartile (EBQ) between well-predicted and ill-predicted individuals for (A) 1D-CNN model and (B) BiLSTM model. Further, within each group we separated correct and incorrect predictions for (C) 1D-CNN and (D) BiLSTM models and compared EBQ. For each box plot, dashed line represents the mean and solid line represents the median. We compared the distributions using Welch's t-test and significant *p*-values (< 0.05) are indicated here.

### 3.4 Feature importance analysis

In our second analysis seeking cognitive underpinnings of model prediction, we conducted a comprehensive analysis of feature importance on 1D-CNN and BiLSTM models to explore the significance of common brain functional networks such as visual, default mode and attention networks. We leveraged functional categorization from the Schafer atlas where each brain region is assigned into one of the several functional networks (ref). In total we included seventeen networks: a) two visual networks, i.e., visual-central and visual-peripheral; b) two somatomotor networks (A & B); c) four attentional networks with two dorsal attention networks and to salient ventral attention networks; d) limbic network; (e) three control networks (A, B, & C); f) three default mode networks (A, B, & C); g) temporal-parietal network; and h) subcortical network. To assess the importance of a given network in model prediction, we permuted all the regions associated with the network and computed the resulting decline in prediction accuracy on the test dataset.

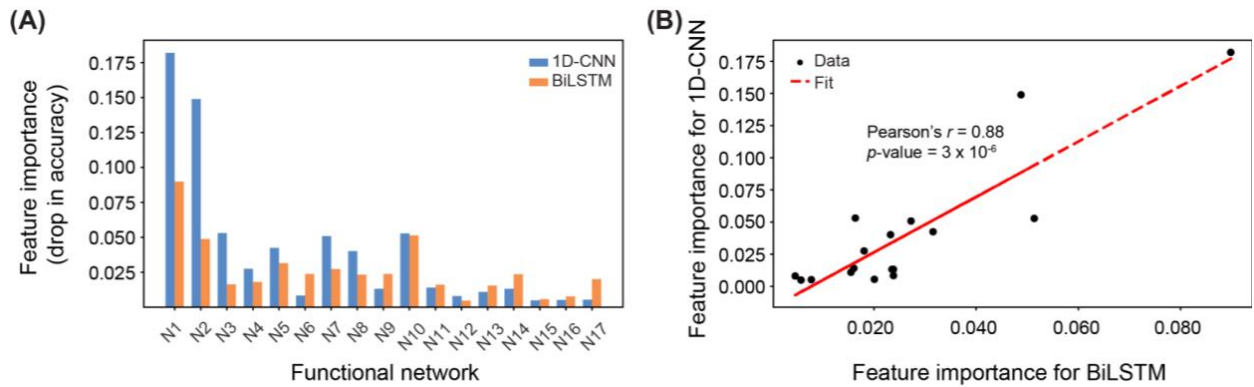

**Figure 8:** Feature importance measured as the mean drop in accuracy resulting from category permutations in the models. We permuted functional networks (i.e., group of regions) derived from the Schaefer atlas (ref) to assess the role of functional networks in model prediction. (A) Feature importance of 1D-CNN and BiLSTM models across different categories (functional networks). Here, N1: Visual-Central, N2: Visual-Peripheral, N3: Somatomotor-A, N4: Somatomoto-B, N5: Dorsal-Attention-A, N6: Dorsal-Attention-B, N7: Salient-Ventral-Attention-A, N8: Salient-Ventral-Attention-B, N9: Limbic, N10: Control-A, N11: Control-B, N12: Control-C, N13: Default Mode Network-A, N14: Default Mode Network-B, N15: Default Mode Network-C, N16: Temporal-Parietal, N17: Subcortical. (B) Scatter plot illustrating the relationship between BiLSTM and CNN model importances, along with a regression line fitted to the data points. Pearson's r and associated p-value is provided to quantify the correlation between the model importances.

### 3.4.1 Impact of cognitive networks on overall model accuracy

As we show in Figure 8, focusing on the overall accuracy for the 1D-CNN model, the visual networks, both central (N1) and peripheral (N2), exhibit high importance values of approximately 0.18 and 0.15, respectively. Following them closely are attention networks (N5, N7, N8) along with Control-A (N10) and Somatomotor-A (N3), each with an importance of around 0.05. Notably, the significance of N1 and N2 features stands out, being three times higher than the others. This suggests their dominance within the 1D-CNN model.

For the BiLSTM model, the feature importance is more evenly distributed with a minimum score of 0.10 across all functional networks. However, similar to the CNN model, visual networks (N1, N2) and Control-A (N10) emerge as the most significant features. This relatively uniform distribution likely results from the reliance on sequential and temporal dependencies encoded within the data. Further, different functional systems may have varied operational scales, which are differentially encoded within the models.

We found a strong positive correlation between the feature importance of the two models (Figure 8B, Pearson's r = 0.88, p-value = 3 x 10$^{-6}$), indicating a meaningful relationship between how features are valued in both the 1D-CNN and BiLSTM models. This suggests that certain features hold consistent importance across different neural network architectures, reinforcing their significance in the predictive performance of the models. These findings were reproduced once we normalized for the numbers of regions in each network (Supplemental Figure S1).

### 3.4.2 Task-specific impact of cognitive networks

Furthermore, we performed a detailed analysis on feature importance by evaluating the drop in the F1-score for each class (task) when different cognitive networks were permuted (Figure 9, Supplemental Figure S2). For 1D-CNN, visual networks (N1, N2) consistently exhibit dominance across all tasks, showcasing relatively higher significance compared to other networks, except for N7: Salient Ventral Attention-A in MOD task which shows similar accuracy drop when permuted, as visual networks. This observation underscores the weight of central and peripheral visual processing in various cognitive tasks.

In the PVT task, the primary importance of networks N1 and N2 is evident, likely indicating the critical role of visual processing in this task, followed by N5: Dorsal-Attention-A, N10: Control-A, N7: Salient-Ventral-Attention-A, N8: Salient-Ventral-Attention-B, and N3: Somatomotor-A, this hierarchy suggests a nuanced interrelation of attentional and motor processing in PVT tasks. Similarly, in the VWM task, the dominance of N1 and N2 underscores the significance of visual processing in maintaining and manipulating visual information over a short period. This is followed by N3: Somatomotor-A and N4: Somatomotor-B further highlights the involvement of somatosensory and motor processing this task performance.

In DOT, the leading networks of N1 and N2, alongside the contributions of N7: Salient-Ventral-Attention-A, N5: Dorsal-Attention-A, N3: Somatomotor-A, and N10: Control-A, suggests the integration of visual, attentional, and motor processes. For MOD, N1, N2 and N7: Salient-Ventral-Attention-A demonstrate similar performance, suggesting a potential overlap or complementarity in their neural processing mechanisms for this particular task. For DYN and Rest, N1 and N2 exhibit considerable dominance across the other networks, suggesting their heightened importance in capturing dynamic and resting-state neural activity patterns. These findings not only highlight the critical role of visual and attentional processes across cognitive tasks but also emphasize the intricate interplay of multiple neural networks in supporting cognitive functions, particularly concerning temporal changes analyzed by the CNN model.

In the BiLSTM model, certain neural networks, particularly N1: Vis-Cent and N2: Vis-Peri, exhibit noteworthy contributions in specific tasks. However, unlike the 1D-CNN model, the distribution of feature importance across other networks appears to be more evenly spread.

Notably, in tasks such as the PVT, N5: DorsAttnA and N10: Cont-A stand out for their elevated feature importance compared to other networks. Similarly, across VWM, DOT, and MOD tasks, N1: Vis-Cent consistently emerges as the most prominent network, followed by N10: Cont-A for DOT and MOD and N9; Limbic for VWM. Furthermore, the DYN, N1: Vis-Cent and N2: Vis-Peri continue to display substantial contributions. However, in resting-state, there is a more equitable distribution of feature importance across networks, indicating a different pattern compared to dynamic tasks.

We further summarize and quantify these findings in Figure 9B-C. In Figure 9B we show the standard deviation of feature importance scores across functional networks within each task, and 1D-CNN shows substantially higher deviation than BiLSTM, indicating higher importance of certain features in 1D-CNN, while more evenly spaced importance scores for a distributed set of functional networks in case of BiLSTM. In Figure 9C we depict the average importance score of functional networks across tasks. The dominance of visual networks (N1 and N2) is clear for the 1D-CNN model. Interestingly, default model, control, and subcortical networks show very little role in model prediction for CNN along with temporal parietal, limbic and a part of dorsal attention network. A more evenly distributed importance score can be visualized for BiLSTM with visual and control-A networks standing out.

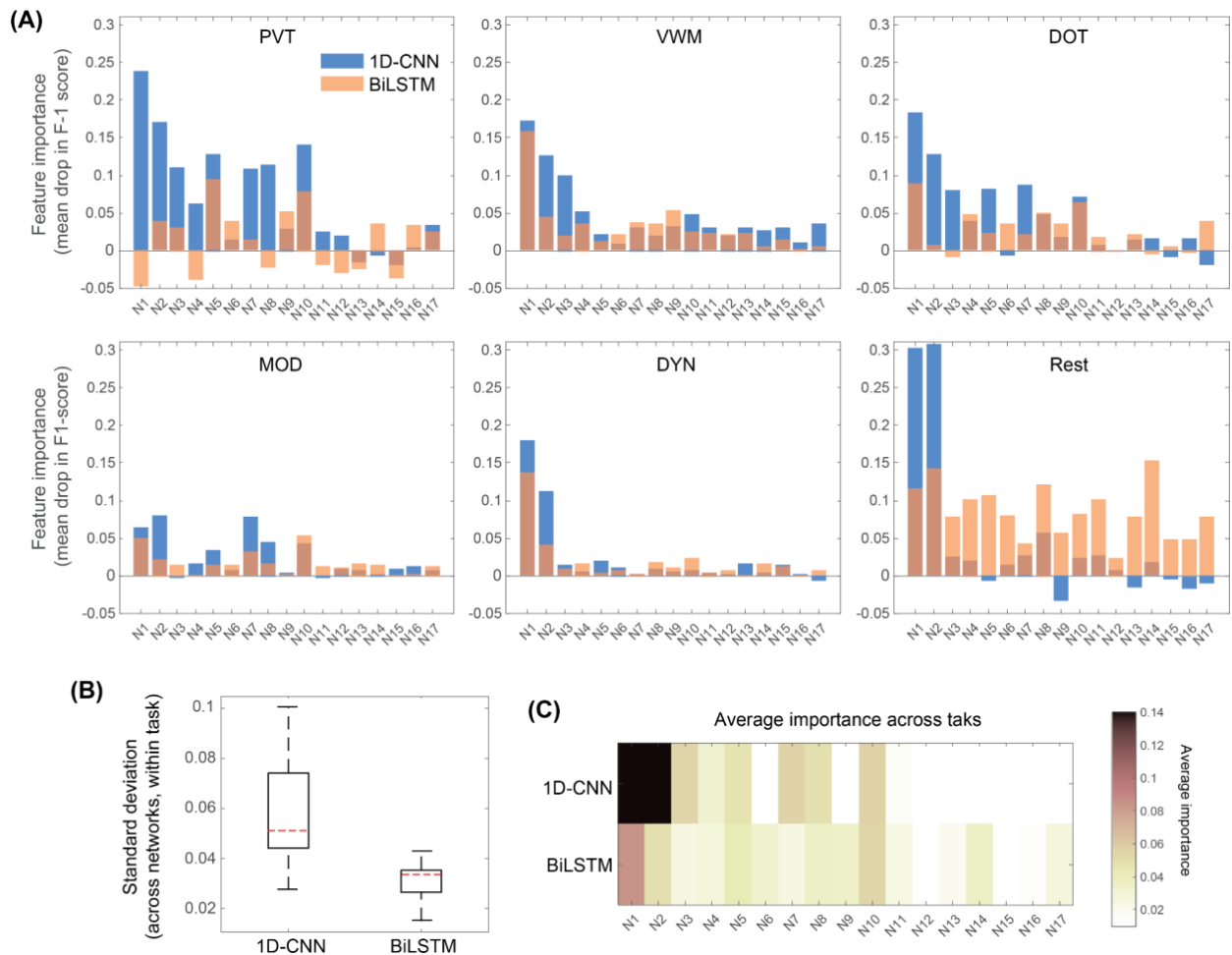

**Figure 9:** Task-specific feature importance. (A) Importance of functional networks in predicting each task as measured by mean drop in F1-score for each task. We utilized functional networks derived from the Schaefer atlas as described in Figure 8. (B) Standard deviation of importance values across networks within each task. (C) Average importance of each network across tasks. Here, N1: Visual-Central, N2: Visual-Peripheral, N3: Somatomotor-A, N4: Somatomoto-B, N5: Dorsal-Attention-A, N6: Dorsal-Attention-B, N7: Salient-Ventral-Attention-A, N8: Salient-Ventral-Attention-B, N9: Limbic, N10: Control-A, N11: Control-B, N12: Control-C, N13: Default Mode Network-A, N14: Default Mode Network-B, N15: Default Mode Network-C, N16: Temporal-Parietal, N17: Subcortical.

## 4. Discussion

To classify different task states using fMRI BOLD signals, we utilized architecturally different 1D-CNN and BiLSTM models, which are well-suited for MRI-based classification and in principle, focus on the distinct aspects of the spatio-temporal signal (Dakka et al., 2017; El Gazzar et al., 2019). CNNs excel at capturing local temporal patterns, this makes them apt for discerning intricate regional patterns of functional brain activity. On the other hand, BiLSTM networks excel at uncovering temporal dependencies within sequential data over longer periods (Qayyum et al., 2022). This capability is especially vital for scrutinizing the dynamic nature of brain activity over time.

Despite the differences in their architecture, our findings point to fundamental similarities in the cognitive underpinnings of cognitive state classification using these two models, improving our understanding, in general, on key components differentiating cognitive states, and opening up new directions regarding the potential mechanisms through which the brain might function under different circumstances. Across both models, we observed high classification accuracies, significantly better than chance, with the 1D-CNN performing slightly better than the BiLSTM. Specifically, states such as DYN, MOD, and VWM showed consistently high classification accuracy across both models. In contrast, states like DOT and PVT had relatively lower classification accuracy. While this can be partly attributed to the smaller sample sizes for DOT and PVT compared to DYN, MOD, and VWM, it is not the only contributing factor. The Rest state, despite having the lowest sample size, achieved classification accuracy closer to that of VWM and MOD. The lower accuracy on PVT and DOT may be due to their varied temporal characteristics, which may improve with higher sample size.

### 4.1 Consistent relationship between behavior performance and prediction accuracy across models

We observed a similar relationship between individual performance and model performance across models. Lower prediction accuracy was associated with low behavioral performance, likely indicating the absence of task specific neural components, essential for accurate classification. Some of our findings indicate trait-based effects, i.e., certain individuals were more likely to have low prediction accuracy, however, these findings were model specific and further work will be needed to understand individual differences in this context. Nevertheless, we found the relationship between low task performance and low prediction accuracy to be robust, despite a relatively crude measure of behavioral performance. Noise in system measurements and differences in model architecture and brain dynamics can significantly impact prediction accuracy too. These factors contribute to variability and highlight the need for more precise behavioral performance measures and additional data to further tease out trait-based features. Addressing these issues in future research will be crucial for improving the robustness and generalizability of cognitive state classification models.

## 4.2 1D-CNN slightly outperformed BiLSTM for overall accuracy

We observed that the 1D-CNN model slightly outperformed the BiLSTM with 3% higher overall accuracy. This performance difference suggests that the CNN architecture might be better suited for this particular dataset, possibly due to its ability to capture local temporal patterns more effectively. CNNs excel at detecting local temporal patterns in the data through convolutional layers, which can be particularly advantageous when dealing with characteristics inherent in fMRI data. The ability of CNNs to perform feature extraction through convolutional filters allows them to discern intricate patterns within the brain's functional activity, contributing to their superior performance in this context.

On the other hand, the slightly lower performance of the BiLSTM model, given our data segmentation, could be attributed to its inherent characteristics. BiLSTMs, while powerful in capturing sequential dependencies and longer temporal dynamics, may struggle with capturing certain complex local temporal patterns that are better handled by CNNs. The bidirectional nature of BiLSTMs allows them to process information from both past and future states, which is beneficial for time-series analysis but may not fully exploit the local temporal patterns present in fMRI data. This limitation could be linked to the model's architecture, which focuses on sequential dependencies with 128 bidirectional units, potentially overlooking fine-grained temporal features. In contrast, the CNN model, with its convolutional layers and filters, excels at detecting intricate local patterns through feature extraction. Despite this, the BiLSTM model demonstrated better sensitivity to individual performance traits than the CNN. The stronger relationship between BiLSTM model predictions and effective behavior quartiles (EBQ) underscores its capability to incorporate individual variability more effectively.

## 4.3 Importance of visual networks in task classification

In our feature importance analysis, we observed a clear dominance of visual networks, indicating that task differences are encoded in how visual information is processed. Another network that stands out across models is N10: Control-A, emphasizing the crucial role of control networks in high level cognitive processing. In case of the 1D-CNN, attention and somatomotor networks also showed high importance, as expected, in executing these tasks that demand maintained attention and button press. Interestingly, we observed a negligible importance of default mode and temporal-parietal networks in task classification, potentially suggesting their more uniform role in supporting cognitive processes.

The observed importance of visual networks may also stem from the inherent architectural and processing differences within the brain. Visual system is a highly connected processing unit, which is sensitive to state-based changes, but probably less so than the other functional regions, such as executive regions and default mode, which need higher functional flexibility to support a wide variety of processing. It is possible that representations of the stimuli within the visual system produce more stable spatio-temporal patterns that are robustly distinguished between states by the models.

Additionally, the feature importance analysis we carried out is constrained by the temporal scale of fMRI signals that are derived from slowly varying hemodynamic responses within the brain. While we find that these responses provide a good basis for state classification, as evident from our high model accuracy, the landscape of feature importance might change with a different measurement, such as EEG, capturing much faster neural responses and brain interactions at a different scale. Further work is required to establish the scale-specificity of neural features in cognitive state classification.

The differential patterns of feature importance distribution between the BiLSTM and 1D-CNN models highlight distinct processing mechanisms and network utilization strategies employed by each model. While the 1D-CNN model may emphasize specific networks more prominently, the BiLSTM model demonstrates a more balanced reliance on a wider range of neural networks, potentially offering advantages in capturing nuanced cognitive processes and dynamic neural activities. Further investigation into the specific mechanisms driving these differences could provide valuable insights into the strengths and limitations of each model architecture.

## 4.4 Do model differences capture functional strategies within the brain?

DNN models are fundamentally inspired by biological neuronal models and stimuli processing, however, the brain uses not just one, but multiple processing strategies. Our findings regarding feature importance might indicate that more localized or specialized processes could function like a CNN, focusing on specialized regional integration, while learning, acquiring new skills, and recovery from brain injury, might require LSTM-like characteristics, relying on a larger distributed set of regions (Bassett et al., 2011). Some existing literature, even though discussed in a different context, indicates this kind of heterogeneity in cognitive processing (Finc et al., 2020; Hu et al., 2018; Wu et al., 2024), however, our assertion is highly speculative and requires assessment with detailed future research.

## 4.5 Methodological considerations

Our study demonstrates the effectiveness of the 1D-CNN and BiLSTM models in classifying fMRI data, but several methodological factors can be considered to improve future studies. Although our sample size was adequate for initial analysis, deep learning models benefit from larger datasets to capture the full variability present in the broader population. Larger and more diverse samples would provide a more comprehensive understanding and enhance the generalizability of our findings. The tasks included in this study are well-established for probing fundamental cognitive processes, yet they may not cover the full spectrum of cognitive functions. Incorporating a wider variety of tasks may provide further insights into cognitive processing within the brain.

The preprocessing steps, such as segmentation and zero-padding, necessary for model training, might introduce biases or affect the interpretability of the results. Our models were not extensively tested for robustness against noise and artifacts, which are common in fMRI data. Evaluating and enhancing model robustness to such disturbances is crucial for reliable real-world applications. While we focused on overall accuracy and common cognitive networks, our study did not delve deeply into potential confounding factors such as age, gender, or baseline cognitive abilities. Future research could explore these aspects to understand how such variability impacts model performance.

The reliance on specific tasks and controlled experimental conditions may limit the generalizability of our findings to more naturalistic settings. Real-world applications often involve more complex and dynamic environments, which our models were not specifically trained or tested for. Despite achieving high accuracy, deep learning models often lack interpretability. Methods to enhance the interpretability of these models, such as layer-wise relevance propagation or attention mechanisms, should be further developed and integrated into the analysis. We attempted to overcome this limitation through feature permutation, which allowed us to assess the importance of different brain regions and improve our understanding of the models' decision-making processes.

## 5. Conclusions

In this study we employed two models: 1D-CNN and BiLSTM, chosen for their complementary ability to capture spatio-temporal features, to classify various cognitive states from fMRI BOLD data. We focused on unraveling the cognitive underpinnings of the classification decisions, and used permutation techniques to calculate feature importance, identifying the most critical brain regions influencing model predictions. Our findings demonstrate a strong relationship between low cognitive performance and poor prediction accuracy across both models. We found the dominance of visual networks in encoding cognitive state differences, however, several considerations emerged in understanding the relative importance of distributed brain regions during state classification, which will need more studies combining deep learning tools with cognitive neuroscience to be fully understood. Further, we found that the 1D-CNN slightly outperformed the BiLSTM in overall accuracy, likely due to its proficiency in transforming high-dimensional temporal information into lower-dimensional vectors through feature extraction. In contrast, the BiLSTM showed better sensitivity to individual performance traits, suggesting its potential for applications requiring detailed behavioral insights. Additionally, the trait-based effects and model-specific differences observed indicate that further research is needed to explore individual variability and model characteristics in greater detail. This study emphasizes the utility of explainable DNN models in understanding neural mechanisms and it provides a solid foundation for future research aimed at advancing the use of deep learning in cognitive neuroscience and enhancing our understanding of brain function.

## Acknowledgements

This research was supported by the U.S. DEVCOM Army Research Laboratory through mission funding (JOG), army educational outreach program contract # W911SR-15-2-0001 (KB), and grant # W911NF2120108 (MK, JB). The views and conclusions contained in this document are those of the authors and should not be interpreted as representing the official policies, either expressed or implied, of the US DEVCOM Army Research Laboratory or the U.S. Government. The U.S. Government is authorized to reproduce and distribute reprints for Government purposes notwithstanding any copyright notation herein.

## References

Avants, B.B., Tustison, N.J., Wu, J., Cook, P.A., Gee, J.C., 2011. An Open Source Multivariate Framework for n-Tissue Segmentation with Evaluation on Public Data. Neuroinform 9, 381–400. https://doi.org/10.1007/s12021-011-9109-y

Bassett, D.S., Wymbs, N.F., Porter, M.A., Mucha, P.J., Carlson, J.M., Grafton, S.T., 2011. Dynamic reconfiguration of human brain networks during learning. Proc. Natl. Acad. Sci. U.S.A. 108, 7641–7646. https://doi.org/10.1073/pnas.1018985108

Breiman, L., 2001. Random Forests. Machine Learning 45, 5–32. https://doi.org/10.1023/A:1010933404324

Chamma, A., Engemann, D.A., Thirion, B., 2023. Statistically Valid Variable Importance Assessment through Conditional Permutations. Advances in Neural Information Processing Systems 36, 67662–67685.

Chyzhyk, D., Savio, A., Graña, M., 2015. Computer aided diagnosis of schizophrenia on resting state fMRI data by ensembles of ELM. Neural Networks 68, 23–33. https://doi.org/10.1016/j.neunet.2015.04.002

Čík, I., Rasamoelina, A.D., Mach, M., Sinčák, P., 2021. Explaining Deep Neural Network using Layer-wise Relevance Propagation and Integrated Gradients, in: 2021 IEEE 19th World Symposium on Applied Machine Intelligence and Informatics (SAMI). Presented at the 2021 IEEE 19th World Symposium on Applied Machine Intelligence and Informatics (SAMI), pp. 000381–000386. https://doi.org/10.1109/SAMI50585.2021.9378686

Covert, I., Lundberg, S., Lee, S.-I., 2021. Explaining by Removing: A Unified Framework for Model Explanation. Journal of Machine Learning Research 22, 1–90.

Dakka, J., Bashivan, P., Gheiratmand, M., Rish, I., Jha, S., Greiner, R., 2017. Learning Neural Markers of Schizophrenia Disorder Using Recurrent Neural Networks. https://doi.org/10.48550/arXiv.1712.00512

Deshpande, G., Masood, J., Huynh, N., Denney, T.S., Dretsch, M.N., 2024. Interpretable Deep Learning for Neuroimaging-Based Diagnostic Classification. IEEE Access 12, 55474–55490. https://doi.org/10.1109/ACCESS.2024.3388911

Deshpande, G., Wang, P., Rangaprakash, D., Wilamowski, B., 2015. Fully Connected Cascade Artificial Neural Network Architecture for Attention Deficit Hyperactivity Disorder Classification From Functional Magnetic Resonance Imaging Data. IEEE Transactions

on Cybernetics 45, 2668–2679. https://doi.org/10.1109/TCYB.2014.2379621

El Gazzar, A., Cerliani, L., van Wingen, G., Thomas, R.M., 2019. Simple 1-D Convolutional Networks for Resting-State fMRI Based Classification in Autism, in: 2019 International Joint Conference on Neural Networks (IJCNN). Presented at the 2019 International Joint Conference on Neural Networks (IJCNN), pp. 1–6. https://doi.org/10.1109/IJCNN.2019.8852002

Finc, K., Bonna, K., He, X., Lydon-Staley, D.M., Kühn, S., Duch, W., Bassett, D.S., 2020. Dynamic reconfiguration of functional brain networks during working memory training. Nat Commun 11, 2435. https://doi.org/10.1038/s41467-020-15631-z

Fintz, M., Osadchy, M., Hertz, U., 2022. Using deep learning to predict human decisions and using cognitive models to explain deep learning models. Sci Rep 12, 4736. https://doi.org/10.1038/s41598-022-08863-0

Gupta, S., Lim, M., Rajapakse, J.C., 2022. Decoding task specific and task general functional architectures of the brain. Hum Brain Mapp 43, 2801–2816. https://doi.org/10.1002/hbm.25817

Hu, J., Du, J., Xu, Q., Yang, F., Zeng, F., Weng, Y., Dai, X., Qi, R., Liu, X., Lu, G., Zhang, Z., 2018. Dynamic Network Analysis Reveals Altered Temporal Variability in Brain Regions after Stroke: A Longitudinal Resting-State fMRI Study. Neural Plasticity 2018, 1–10. https://doi.org/10.1155/2018/9394156

Janitza, S., Celik, E., Boulesteix, A.-L., 2018. A computationally fast variable importance test for random forests for high-dimensional data. Adv Data Anal Classif 12, 885–915. https://doi.org/10.1007/s11634-016-0276-4

Li, G., Li, F., Xu, C., Fang, X., 2022. A spatial-temporal layer-wise relevance propagation method for improving interpretability and prediction accuracy of LSTM building energy prediction. Energy and Buildings 271, 112317. https://doi.org/10.1016/j.enbuild.2022.112317

Li, H., Tian, Y., Mueller, K., Chen, X., 2019. Beyond saliency: Understanding convolutional neural networks from saliency prediction on layer-wise relevance propagation. Image and Vision Computing 83–84, 70–86. https://doi.org/10.1016/j.imavis.2019.02.005

Makris, N., Goldstein, J.M., Kennedy, D., Hodge, S.M., Caviness, V.S., Faraone, S.V., Tsuang, M.T., Seidman, L.J., 2006. Decreased volume of left and total anterior insular lobule in schizophrenia. Schizophrenia Research 83, 155–171. https://doi.org/10.1016/j.schres.2005.11.020

Meng, L., Ge, K., 2022. Decoding Visual fMRI Stimuli from Human Brain Based on Graph Convolutional Neural Network. Brain Sciences 12, 1394. https://doi.org/10.3390/brainsci12101394

Meszlényi, R.J., Buza, K., Vidnyánszky, Z., 2017. Resting State fMRI Functional Connectivity-Based Classification Using a Convolutional Neural Network Architecture. Front. Neuroinform. 11. https://doi.org/10.3389/fninf.2017.00061

Nakuci, J., Wasylyshyn, N., Cieslak, M., Elliott, J.C., Bansal, K., Giesbrecht, B., Grafton, S.T., Vettel, J.M., Garcia, J.O., Muldoon, S.F., 2023. Within-subject reproducibility varies in multi-modal, longitudinal brain networks. Sci Rep 13, 6699. https://doi.org/10.1038/s41598-023-33441-3

Patel, P., Aggarwal, P., Gupta, A., 2016. Classification of Schizophrenia versus normal subjects using deep learning, in: Proceedings of the Tenth Indian Conference on Computer Vision, Graphics and Image Processing, ICVGIP '16. Association for Computing Machinery, New York, NY, USA, pp. 1–6. https://doi.org/10.1145/3009977.3010050

Qayyum, A., Ahamed Khan, M.K.A., Benzinou, A., Mazher, M., Ramasamy, M., Aramugam, K., Deisy, C., Sridevi, S., Suresh, M., 2022. An Efficient 1DCNN–LSTM Deep Learning Model for Assessment and Classification of fMRI-Based Autism Spectrum Disorder, in: Raj, J.S., Kamel, K., Lafata, P. (Eds.), Innovative Data Communication Technologies

and Application. Springer Nature, Singapore, pp. 1039–1048. https://doi.org/10.1007/978-981-16-7167-8_77

Quaak, M., van de Mortel, L., Thomas, R.M., van Wingen, G., 2021. Deep learning applications for the classification of psychiatric disorders using neuroimaging data: Systematic review and meta-analysis. NeuroImage: Clinical 30, 102584. https://doi.org/10.1016/j.nicl.2021.102584

Qureshi, M.N.I., Oh, J., Lee, B., 2019. 3D-CNN based discrimination of schizophrenia using resting-state fMRI. Artificial Intelligence in Medicine 98, 10–17. https://doi.org/10.1016/j.artmed.2019.06.003

Rungratsameetaweemana, N., Lainscsek, C., Cash, S.S., Garcia, J.O., Sejnowski, T.J., Bansal, K., 2022. Brain network dynamics codify heterogeneity in seizure evolution. Brain Commun 4, fcac234. https://doi.org/10.1093/braincomms/fcac234

Schaefer, A., Kong, R., Gordon, E.M., Laumann, T.O., Zuo, X.-N., Holmes, A.J., Eickhoff, S.B., Yeo, B.T.T., 2018. Local-Global Parcellation of the Human Cerebral Cortex from Intrinsic Functional Connectivity MRI. Cereb Cortex 28, 3095–3114. https://doi.org/10.1093/cercor/bhx179

Schuster, M., Paliwal, K.K., 1997. Bidirectional recurrent neural networks. IEEE Transactions on Signal Processing 45, 2673–2681. https://doi.org/10.1109/78.650093

Segal, A., Parkes, L., Aquino, K., Kia, S.M., Wolfers, T., Franke, B., Hoogman, M., Beckmann, C.F., Westlye, L.T., Andreassen, O.A., Zalesky, A., Harrison, B.J., Davey, C.G., Soriano-Mas, C., Cardoner, N., Tiego, J., Yücel, M., Braganza, L., Suo, C., Berk, M., Cotton, S., Bellgrove, M.A., Marquand, A.F., Fornito, A., 2023. Regional, circuit and network heterogeneity of brain abnormalities in psychiatric disorders. Nat Neurosci 26, 1613–1629. https://doi.org/10.1038/s41593-023-01404-6

Shermadurai, P., Thiyagarajan, K., 2023. Deep Learning Framework for Classification of Mental Stress from Multimodal Datasets. RIA 37, 155–163. https://doi.org/10.18280/ria.370119

Srinivasagopalan, S., Barry, J., Gurupur, V., Thankachan, S., 2019. A deep learning approach for diagnosing schizophrenic patients. Journal of Experimental & Theoretical Artificial Intelligence 31, 803–816. https://doi.org/10.1080/0952813X.2018.1563636

Thomas, A.W., Ré, C., Poldrack, R.A., 2022. Interpreting mental state decoding with deep learning models. Trends in Cognitive Sciences 26, 972–986. https://doi.org/10.1016/j.tics.2022.07.003

Wang, C., Xiao, Z., Wang, B., Wu, J., 2019. Identification of Autism Based on SVM-RFE and Stacked Sparse Auto-Encoder. IEEE Access 7, 118030–118036. https://doi.org/10.1109/ACCESS.2019.2936639

Wu, K., Jelfs, B., Neville, K., Mahmoud, S.S., He, W., Fang, Q., 2024. Dynamic Reconfiguration of Brain Functional Network in Stroke. IEEE Journal of Biomedical and Health Informatics 28, 3649–3659. https://doi.org/10.1109/JBHI.2024.3371097

# Supplemental Material

# Cognitive Networks and Performance Drive fMRI-Based State Classification Using DNN Models

Murat Kucukosmanoglu[1], Javier O. Garcia[2], Justin Brooks[1,3,4], Kanika Bansal[2,3,*]

[1] D-Prime LLC, McLean, VA 22101 USA
[2] Humans in Complex Systems, US DEVCOM Army Research Laboratory, Aberdeen Proving Ground, MD 21005 USA
[3] Department of Computer Science and Electrical Engineering, University of Maryland, Baltimore County, Baltimore, MD 21250 USA
[4] Tanzen Medical Inc., Baltimore, MD, USA

* Corresponding author email: phy.kanika@gmail.com

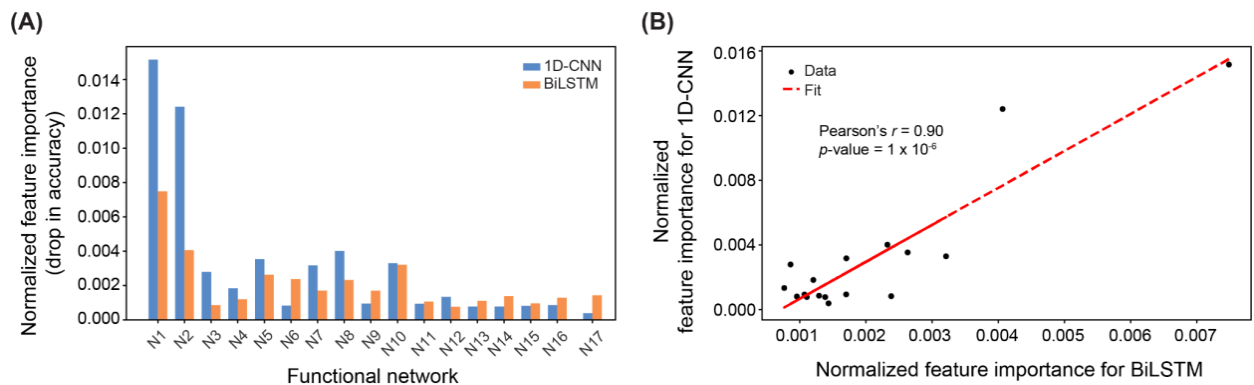

**Supplementary Figure S1:** Feature importance, similar to Figure 8, except that the importance scores are normalized by the number of brain regions in each functional network. (A) Normalized feature importance of 1D-CNN and BiLSTM models across different categories (functional networks). Here, N1: Visual-Central, N2: Visual-Peripheral, N3: Somatomotor-A, N4: Somatomoto-B, N5: Dorsal-Attention-A, N6: Dorsal-Attention-B, N7: Salient-Ventral-Attention-A, N8: Salient-Ventral-Attention-B, N9: Limbic, N10: Control-A, N11: Control-B, N12: Control-C, N13: Default Mode Network-A, N14: Default Mode Network-B, N15: Default Mode Network-C, N16: Temporal-Parietal, N17: Subcortical. (B) Scatter plot illustrating the relationship between BiLSTM and CNN model importances, along with a regression line fitted to the data points. Pearson's r and associated p-value is provided to quantify the correlation between the model importances.

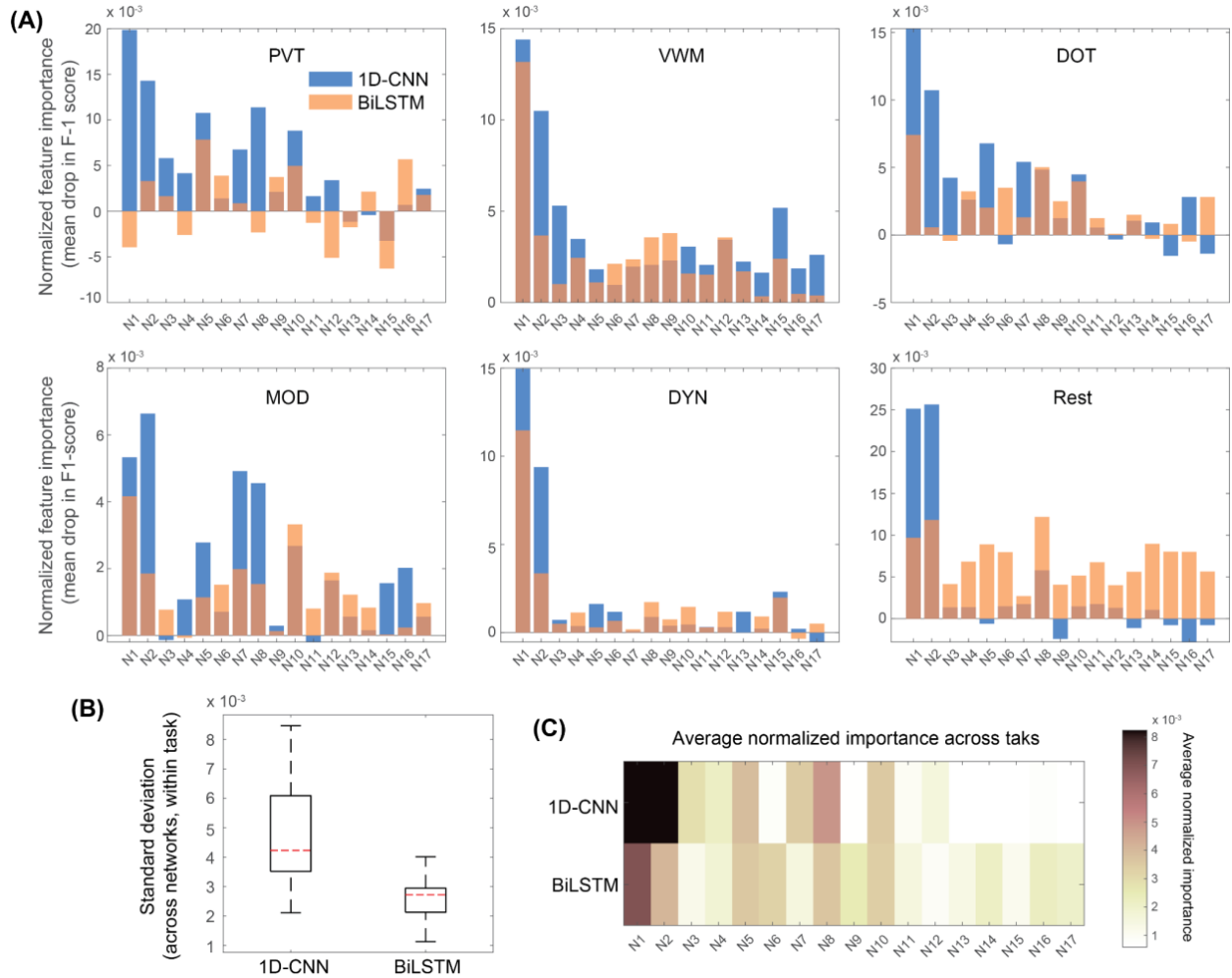

**Supplementary Figure S2:** Task-specific feature importance, similar to Figure 9, except that the importance scores are normalized by the number of brain regions in each functional network. (A) Normalized importance of functional networks in predicting each task. (B) Standard deviation of normalized importance values across networks within each task. (C) Average normalized importance of each network across tasks. Here, N1: Visual-Central, N2: Visual-Peripheral, N3: Somatomotor-A, N4: Somatomoto-B, N5: Dorsal-Attention-A, N6: Dorsal-Attention-B, N7: Salient-Ventral-Attention-A, N8: Salient-Ventral-Attention-B, N9: Limbic, N10: Control-A, N11: Control-B, N12: Control-C, N13: Default Mode Network-A, N14: Default Mode Network-B, N15: Default Mode Network-C, N16: Temporal-Parietal, N17: Subcortical.